\documentclass[12pt]{article}
\usepackage{graphicx,rotating,epsfig,amsmath,amssymb,color,multirow,pstricks,pst-plot}
\newlength{\digitwidth} 
\newcommand{\cinst}[2]{$^{\mathrm{#1)}}$~#2\par}
\newcommand{\crefi}[1]{$^{\mathrm{#1)}}$}
\newcommand{\HRule}{\rule{0.4\linewidth}{0.3mm}}
\newcommand{\mlab}[1]%
    {\mbox{}\marginpar{\raggedright\hspace{0pt}\footnotesize #1}}

\newcommand\epjc[3]  {
        {{\it Eur.\ Phys.\ J. }{\bf C #1} (#2) #3}}

\newcommand\npb[3]   {
        {{\it Nucl.\ Phys.\ }{\bf B #1} (#2) #3}}
\newcommand\plb[3]   {
        {{\it Phys.\ Lett.\ }{\bf B #1} (#2) #3}}
\newcommand\prc[3]   {
        {{\it Phys.\ Rev.\ }{\bf C #1} (#2) #3}}
\newcommand\prd[3]   {
        {{\it Phys.\ Rev.\ }{\bf D #1} (#2) #3}}
\newcommand\prl[3]   {
        {{\it Phys.\ Rev.\ Lett.\ }{\bf #1} (#2) #3}}
\newcommand\zpc[3]   {
        {{\it Z.\ Physik }{\bf C #1} (#2) #3}}

\begin{document}

\begingroup
\thispagestyle{empty} \baselineskip=14pt
\parskip 0pt plus 5pt

\begin{center}
{\large EUROPEAN LABORATORY FOR PARTICLE PHYSICS}
\end{center}

\bigskip
\begin{flushright}
CERN--PH--EP\,/\,2008--xxx\\
August 25, 2008
\end{flushright}

\bigskip
\begin{center}
{\Large\bf \boldmath Study of $\psi^\prime$ and $\chi_c$ decays as feed-down sources \\[0.3 cm]
of J$/\psi$ hadro-production}

\bigskip\bigskip

Pietro Faccioli\crefi{1},
Carlos Louren\c{c}o\crefi{2},
Jo\~ao Seixas\crefi{1,3}
and Hermine K.\ W\"ohri\crefi{1}

\bigskip\bigskip\bigskip
\textbf{Abstract}

\end{center}

\begingroup
\leftskip=0.4cm \rightskip=0.4cm
\parindent=0.pt

  The interpretation of the J$/\psi$ suppression patterns
  observed in nuclear collisions, at CERN and RHIC, as a signature of
  the formation of a deconfined phase of QCD matter, requires knowing
  which fractions of the measured J$/\psi$ yields, in pp collisions,
  are due to decays of heavier charmonium states.  From a detailed
  analysis of the available mid-rapidity charmonium hadro-production
  cross sections, or their ratios, we determine that the J$/\psi$
  feed-down contributions from $\psi^\prime$ and $\chi_c$ decays are,
  respectively, $(8.1 \pm 0.3) \%$ and $(25 \pm 5) \%$.  These
  proton-proton values are derived from global averages of the
  proton-nucleus measurements, assuming that the charmonium states are
  exponentially absorbed with the length of matter they traverse in
  the nuclear targets.

\bigskip



\endgroup

\begin{center}
\end{center}

\vfill
\begin{flushleft}
\HRule\\

\cinst{1} {Laborat\'orio de Instrumenta\c{c}\~ao e F\'{\i}sica Experimental de
  Part\'{\i}culas (LIP),\\ ~~~Lisbon, Portugal}
\cinst{2} {CERN, Geneva, Switzerland}
\cinst{3} {Instituto Superior T\'ecnico (IST) and Centro de  F\'{\i}sica Te\'orica de
  Part\'{\i}culas\\ ~~~(CFTP), Lisbon, Portugal}
\end{flushleft}
\endgroup

\newpage
\thispagestyle{empty} ~ 

\newpage
\pagenumbering{arabic} \setcounter{page}{1} \sloppy

\section{Introduction and motivation}

In the very hot and dense, strongly interacting matter produced in
high-energy nuclear collisions, it is expected that the QCD binding
potential is screened, the screening level increasing with the energy
density of the created system~\cite{bib:Karsch}. Depending on the
screening level, it may happen that the charmonium states
``dissolve'' into open charm mesons~\cite{bib:MatsuiSatzKarschMehr}.
Since different quarkonium states have different binding energies,
they are expected to dissolve at successive ``thresholds'' in the
energy density or temperature of the
medium~\cite{bib:KharzeevKarschSatz}.  In particular, the
$\psi^\prime$ and $\chi_c$ states should be easier to ``melt'' than
the more strongly bound J/$\psi$ state. Therefore, a ``spectral
analysis'' of the charmonium production yields, in several collision
systems (from light to heavy nuclei) and in several collision
centralities (from peripheral to central), should provide very
interesting information concerning the nature of the produced matter.

Experimentally, it has not yet been possible to directly measure the
production yields of the $\chi_c$ state in heavy-ion collisions.
However, it is well known that a significant fraction of the J/$\psi$
mesons observed in pp collisions are, in fact, produced by $\chi_c$
radiative decays. The J/$\psi$ production yield measured in heavy-ion
collisions could thus show a significant level of suppression, even
if the collision system under scrutiny has not reached high enough
energy densities to melt the directly produced J/$\psi$ state.  In
particular, it could very well be that the J/$\psi$ suppression
pattern measured at the SPS and RHIC is essentially due to the melting
of the $\psi^\prime$ and $\chi_c$
states~\cite{bib:KharzeevKarschSatz}.

The picture is made more complex by the fact that already in
proton-nucleus collisions the charmonium production cross sections
scale less than linearly with the number of binary nucleon-nucleon
collisions. As we will recall in this paper, this ``normal nuclear
absorption'' has been seen (at the CERN-SPS and at Fermilab) to be
significantly stronger for $\psi^\prime$ mesons than for J/$\psi$
mesons. While the existing $\chi_c$ data are much less accurate, there
is no reason to assume that the $\chi_c$ and J/$\psi$ mesons have
the same ``nuclear dependence''. A stronger $\chi_c$ ``normal nuclear
absorption'' would decrease the yield of J/$\psi$ mesons produced from
$\chi_c$ decays and, hence, would account for part of the ``anomalous
J/$\psi$ suppression'' seen in heavy-ion collisions.  How much of that
``anomaly'' might be due to the normal nuclear absorption of the
$\psi^\prime$ and $\chi_c$ mesons depends on the fractions of J/$\psi$
mesons produced by $\psi^\prime$ and $\chi_c$ decays.  These
considerations underline the importance of knowing these fractions, in
elementary collisions, as accurately as possible.

In the existing literature, the feed-down fractions are generally
assumed to be around $10\,\%$ for the $\psi^\prime$ and around $30$ or
$40\,\%$ for the $\chi_c$, usually without mentioning
experimental measurements or their uncertainties.  Yet, the
J/$\psi$ feed-down fraction from $\psi^\prime$ decays can be rather
precisely determined, from data collected by SPS and Fermilab
experiments.  The $\chi_c$ case has been much less investigated but
recent measurements, by the HERA-B experiment, indicate a J/$\psi$
feed-down fraction from $\chi_c$ decays of around $20\,\%$, considerably
lower than the previously assumed values.

This paper presents a quantitative analysis of the presently available
data on feed-down contributions to J/$\psi$ hadro-production, at fixed
target energies.  The relevant measurements are presented and reviewed
in Section~\ref{sec:measurements}.  New ``world averages'' (including
uncertainties) of the J/$\psi$ feed-down fractions from $\psi^\prime$
and $\chi_c$ decays are then derived and discussed, in
Sections~\ref{sec:psiprime} and~\ref{sec:chic}.

\section{Overview of available measurements}
\label{sec:measurements}

In this section we briefly review the existing measurements of
$\psi^\prime$ and $\chi_c$ hadro-production, which can be used to
constrain the corresponding fractions of indirectly produced
J$/\psi$'s. These fractions are defined with respect to the total
(inclusive) J$/\psi$ yield:
\begin{equation}
R(\psi^\prime) = \frac{N(\mathrm{J}/\psi \; \mathrm{from} \;
\psi^\prime)}{N_{\mathrm{incl}}(\mathrm{J}/\psi)} = \frac{\sigma(\psi^\prime) \cdot
B(\psi^\prime \rightarrow \mathrm{J}/\psi X)} {\sigma(\mathrm{J}/\psi)} \quad,
\end{equation}
\noindent and analogous for $R(\chi_c)$.

The $\psi^\prime$ mainly decays into a J$/\psi$ and a pair of pions.
However, most experiments measure the J$/\psi$ and $\psi^\prime$
dilepton decays, reporting results for the yield ratio
\begin{equation}
\rho(\psi^\prime) = \frac{\sigma(\psi^\prime) \cdot B(\psi^\prime \rightarrow l^+
l^-)}{\sigma(\mathrm{J}/\psi) \cdot B(\mathrm{J}/\psi \rightarrow l^+ l^-)}\quad.
\end{equation}
This quantity is directly related to the $\psi^\prime$-to-J$/\psi$
feed-down fraction, $R(\psi^\prime)$, through a simple combination of
branching ratios,
\begin{equation}
R(\psi^\prime) = \left[~\frac{B(\mathrm{J}/\psi \rightarrow l^+ l^-)}{B(\psi^\prime
\rightarrow l^+ l^-)} ~ B(\psi^\prime \rightarrow \mathrm{J}/\psi~ X) ~\right] ~
\rho(\psi^\prime) = (4.53 \pm 0.13) ~ \rho(\psi^\prime)\quad,
\end{equation}
where the numerical values were derived from the PDG tables~\cite{bib:PDG}.

The $R(\chi_c)$ values are obtained dividing the number of J$/\psi$'s
resulting from the radiative decays $\chi_{c} \rightarrow
\mathrm{J}/\psi~ \gamma$ by the total number of observed J$/\psi$'s.

\bigskip

The running conditions of the experiments providing these measurements
are summarized in Tables~\ref{tab:Rpsip} and~\ref{tab:Rchic}. Most
experiments made use of proton or pion beams of different energies
incident on several target nuclei, but collider experiments have also
provided some results.  The different detectors covered
$x_{\mathrm{F}}$ intervals extending from slightly backward
to very forward values.  Average $x_{\mathrm{F}}$ values have been
estimated for each experiment, either from the measured distributions
or from the variation of the statistical errors in the
efficiency-corrected spectra.

\begin{table}[h!]
\begin{center}
\begin{tabular}{|l|c|c|c|c|c|}
    \hline
    \footnotesize{Experiment}  &
    \footnotesize{Collision system} &
    \footnotesize{$E_{\mathrm{beam}}$ [GeV]} &
    \footnotesize{Phase space} & \footnotesize{$\langle x_{\mathrm{F}} \rangle$}  \\

    \hline
    \hline
    E331
    \cite{bib:psip_E331} &
    p-C  & $225$ & $0 < x_{\mathrm{F}} < 0.7$ & $\simeq 0.3$  \\

    \hline
    E444
    \cite{bib:psip_E444} &
    p-C  & $225$ & $0 < x_{\mathrm{F}} < 0.9$ & $\simeq 0.35$  \\

    \hline
    E705
    \cite{bib:psip_E705} &
    p-Li  & $300$ & $-0.1 < x_{\mathrm{F}} < 0.5$ & $\simeq 0.2$  \\

    \hline
    E288
    \cite{bib:psip_E288} &
    p-Be  & $400$ & $-0.6 < x_{\mathrm{F}} < 0.8$ & $\simeq 0.1$  \\

    \hline
    NA38
    \cite{bib:psip_NA38} &
    \begin{tabular}{@{}c@{}} p-W/U \\ p-C/Al/Cu/W \\  \end{tabular} &
    \begin{tabular}{@{}c@{}} $200$ \\ $450$ \\  \end{tabular} &
    $-0.4 < y_{\mathrm{cm}} < 0.6$ & $\simeq 0$  \\

    \hline
    NA51
    \cite{bib:psip_NA51} &
    p-H/D &
    $450$ &
    $-0.4 < y_{\mathrm{cm}} < 0.6$ & $\simeq 0$  \\

    \hline
    NA50 96/98
    \cite{bib:psip_NA50_450GeV} &
    \begin{tabular}{@{}c@{}} p-Be/Al/Cu/ \\ Ag/W \\  \end{tabular} &
    $450$ &
    $-0.5 < y_{\mathrm{cm}} < 0.5$ & $\simeq 0$  \\

    \hline
    NA50 2000
    \cite{bib:psip_NA50_400GeV} &
    \begin{tabular}{@{}c@{}} p-Be/Al/Cu/ \\ Ag/W/Pb \\  \end{tabular} &
    $400$ &
    $-0.425 < y_{\mathrm{cm}} < 0.575$ & $\simeq 0$  \\

    \hline
    E771
    \cite{bib:psip_E771} &
    p-Si  & $800$ & $-0.05 < x_{\mathrm{F}} < 0.25$ & $\simeq 0.1$  \\

    \hline
    E789
    \cite{bib:psip_E789} &
    p-Au  & $800$ & $-0.03 < x_{\mathrm{F}} < 0.15$ & $\simeq 0.06$  \\

    \hline
    E866
    \cite{bib:psip_E866} &
    p-Be/Fe/W &
    $800$ &
    $-0.1 < x_{\mathrm{F}} < 0.8$ & $\simeq 0.3$  \\

    \hline
    HERA-B
    \cite{bib:psip_HERA-B} &
    p-C/Ti/W &
    $920$ &
    $-0.35 < x_{\mathrm{F}} < 0.1$ & $-0.065$  \\

    \hline
    \hline
    WA39
    \cite{bib:psip_WA39} &
    $\pi^{\pm}$-W  & $39.5$ & $-0.5 < x_{\mathrm{F}} < 0.8$ & $\simeq 0.2$  \\

    \hline
    E537
    \cite{bib:psip_E537} &
    $\pi^{-}$-W  & $125$ & $0 < x_{\mathrm{F}} < 1$ & $\simeq 0.3$  \\

    \hline
    WA11
    \cite{bib:psip_WA11} &
    $\pi^{-}$-Be  & $150$ & $-0.4 < x_{\mathrm{F}} < 0.9$ & $\simeq 0.3$  \\

    \hline
    E331
    \cite{bib:psip_E331} &
    $\pi^{+}$-C  & $225$ & $0 < x_{\mathrm{F}} < 0.9$ & $\simeq 0.35$  \\

    \hline
    E444
    \cite{bib:psip_E444} &
    $\pi^{\pm}$-C  & $225$ & $0 < x_{\mathrm{F}} < 1$ & $\simeq 0.4$  \\

    \hline
    E615
    \cite{bib:psip_E615} &
    $\pi^{-}$-W  & $253$ & $0.3 < x_{\mathrm{F}} < 1$ & $\simeq 0.6$  \\

    \hline
    E705
    \cite{bib:psip_E705} &
    $\pi^{\pm}$-Li  & $300$ & $-0.1 < x_{\mathrm{F}} < 0.5$ & $\simeq 0.2$  \\

    \hline
    E672-706
    \cite{bib:psip_E672} &
    $\pi^{-}$-Be  & $515$ & $0.1 < x_{\mathrm{F}} < 0.8$ & $\simeq 0.4$  \\

    \hline
    \hline
    \footnotesize{Experiment}  &
    \footnotesize{Collision system} &
    \footnotesize{$\sqrt{s}$ [GeV]} &
    \footnotesize{Phase space} & \footnotesize{$\langle x_{\mathrm{F}} \rangle$}  \\

    \hline
    \hline
    ISR
    \cite{bib:psip_ISR} &
    pp &
    $58$ \footnotesize{(avg.)} &
    $y_{\mathrm{cm}} \simeq 0$ & $0$  \\

    \hline
\end{tabular}
\caption{\label{tab:Rpsip} Global features characterizing the existing measurements of
the $\psi^\prime$-to-J$/\psi$ cross-section ratio in proton-nucleus, pion-nucleus and
proton-proton collisions.}
\end{center}
\end{table}


\begin{table}[h!]
\begin{center}
\begin{tabular}{|l|c|c|c|c|c|}
    \hline
    \footnotesize{Experiment}  &
    \footnotesize{Collision system} &
    \footnotesize{$E_{\mathrm{beam}}$ [GeV]} &
    \footnotesize{Phase space} & \footnotesize{$\langle x_{\mathrm{F}} \rangle$}  \\

    \hline
    \hline
    E369-610-673
    \cite{bib:chic_E369} &
    p-Be &
    $225$ \footnotesize{(avg.)}  &
    $0.1 < x_{\mathrm{F}} < 0.6$ & $0.32$  \\

    \hline
    E705
    \cite{bib:chic_E705} &
    p-Li  & $300$ & $-0.1 < x_{\mathrm{F}} < 0.5$ & $\simeq 0.2$  \\

    \hline
    E771
    \cite{bib:chic_E771} &
    p-Si  & $800$ & $-0.05 < x_{\mathrm{F}} < 0.25$ & $\simeq 0.1$  \\

    \hline
    HERA-B 2000
    \cite{bib:chic_HERA-B_2000} &
    p-C/Ti &
    $920$ & $-0.25 < x_{\mathrm{F}} < 0.15$ & $-0.035$  \\

    \hline
    HERA-B 2003
    \cite{bib:chic_HERA-B_2003} &
    p-C/W &
    $920$ & $-0.35 < x_{\mathrm{F}} < 0.15$ & $-0.065$  \\

    \hline
    \hline
    SERPUKHOV-140
    \cite{bib:chic_SERPUKHOV-140} &
    $\pi^{-}$-H & $38$ & $0.3 < x_{\mathrm{F}} < 0.8$ & $\simeq 0.5$  \\

    \hline
    WA11
    \cite{bib:chic_WA11} &
    $\pi^{-}$-Be & $185$ & $-0.4 < x_{\mathrm{F}} < 0.9$ & $\simeq 0.3$  \\

    \hline
    E369-610-673
    \cite{bib:chic_E369} &
    $\pi^{-}$-Be \footnotesize{(mostly)} &
    $209$ \footnotesize{(avg.)}  &
    $0 < x_{\mathrm{F}} < 0.8$ & $0.43$  \\

    \hline
    E705
    \cite{bib:chic_E705} &
    $\pi^{\pm}$-Li  & $300$ & $-0.1 < x_{\mathrm{F}} < 0.5$ & $\simeq 0.2$  \\

    \hline
    E672-706
    \cite{bib:chic_E672} &
    $\pi^{-}$-Be  & $515$ & $0.1 < x_{\mathrm{F}} < 0.8$ & $\simeq 0.4$  \\

    \hline
    \hline
    \footnotesize{Experiment}  &
    \footnotesize{Collision system} &
    \footnotesize{$\sqrt{s}$ [GeV]} &
    \footnotesize{Phase space} & \footnotesize{$\langle x_{\mathrm{F}} \rangle$}  \\

    \hline
    \hline
    ISR
    \cite{bib:chic_ISR} &
    pp &
    $58$ \footnotesize{(avg.)}  &
    $y_{\mathrm{cm}} \simeq 0$ & $0$  \\

    \hline
    CDF
    \cite{bib:chic_CDF} &
    p$\bar{\mathrm{p}}$  &
    $1800$ &
    $|y_{\mathrm{cm}}| < 0.6$ & $0$  \\

    \hline
\end{tabular}
\caption{\label{tab:Rchic} Global features characterizing the existing measurements of
the $R(\chi_c)$ feed-down ratio in proton-nucleus, pion-nucleus and proton-(anti)proton
collisions.}
\end{center}
\end{table}


While all the relevant experiments are listed in these tables, only a
restricted subsample of data is used in the present analysis. Most
importantly, we do not use results obtained on the basis of forward
$x_{\mathrm{F}}$ data. In fact, it is well established (in particular
by E866~\cite{bib:psip_E866}, for $x_{\mathrm{F}} > 0.2$) that the
J$/\psi$ and $\psi^\prime$ production cross sections measured at high
$x_{\mathrm{F}}$, in nuclear targets, exhibit a much stronger nuclear
absorption than the corresponding mid-rapidity values.  A significant
role in this behaviour should be played by nuclear effects on the
parton distribution functions of the target nucleons and by the energy
loss of the beam partons (or of the produced state) traversing the
nuclear matter.  Other effects may also contribute, such as intrinsic
heavy-quark components of the scattering nucleons and interactions
with other produced hadrons (``comovers''), as discussed in
Ref.~\cite{bib:VogtXF}.  The data at high $x_{\mathrm{F}}$ may,
therefore, reflect a non-trivial cocktail of production and absorption
mechanisms, certainly not easy to disentangle and quantify. For this
reason, in this paper we concentrate on the analysis of the
mid-rapidity data, a choice consistent with our goal of determining
reference values for the interpretation of the existing observations
of quarkonium suppression in nucleus-nucleus collisions, also made at
mid-rapidity.  This means, in particular, that the data obtained in
pion-nucleus collisions, always significantly extending towards high
$x_{\mathrm{F}}$ values, are left out from the present analysis.
Further details on the data selection are discussed in the following
paragraphs. A broader analysis of the existing measurements,
trying to take into account possible kinematic dependencies induced by
nuclear effects, will be the topic of a future investigation.

The current experimental knowledge concerning $\psi^\prime$ production
in proton-nucleus collisions is essentially determined by the accurate
measurements performed by NA50/NA51 at the CERN-SPS and by E866 at
Fermilab, using several target nuclei.  Thanks to their several
million reconstructed J$/\psi$ events, these measurements provide
precise determinations of the $\psi^\prime$-to-J$/\psi$ cross-section
ratio and of its nuclear dependence.  Given their much lower level of
precision, the other existing measurements (obtained in several
kinematical windows and using a single target nucleus) have no
influence on a global average and were, therefore, left out of the
present analysis.

The NA50 data sets were obtained at $450$~\cite{bib:psip_NA50_450GeV}
and $400$~GeV~\cite{bib:psip_NA50_400GeV}, respectively with five and
six targets.  Two statistically independent $450$~GeV data samples
were collected, at different proton beam intensities.  In our study we
used their average, after considering the corrections reported in
Ref.~\cite{bib:psip_NA50_400GeV}.

The J$/\psi$ and $\psi^\prime$ results of E866~\cite{bib:psip_E866}
were reported as ratios between the yields obtained with heavy and
light targets (W/Be and Fe/Be), as a function of $x_{\mathrm{F}}$.
These measurements provide heavy-over-light \emph{ratios} of
$\rho(\psi^\prime)$, which cannot help determining the feed-down
fraction value but constrain the difference between the nuclear
absorption rates of the two charmonium states. Although data points
exist up to $x_{\mathrm{F}} = 0.8$, we restricted our study to the
range $x_{\mathrm{F}} < 0.2$, where only the W/Be ratio was reported.

Also the $\chi_c$ measurements mentioned in Table~\ref{tab:Rchic}
deserve a few remarks.  The error we quote for the E705 value
reflects the systematic uncertainties mentioned in
their paper~\cite{bib:chic_E705}.  Although the E771
publication~\cite{bib:chic_E771} provides no explicit value for
$R(\chi_c)$, it should be possible to derive it from the quoted
$\chi_{c1}$ and $\chi_{c2}$ cross sections or, alternatively, from the
yields of reconstructed $\chi_{c1}$, $\chi_{c2}$ and J$/\psi$ mesons,
using the quoted efficiencies.  It turns out, however, that these two
methods lead to significantly different results.  Moreover, the
information provided is insufficient to properly evaluate the
$R(\chi_c)$ uncertainty.  Therefore, we did not consider the E771
measurement in our analysis.

It is worth noting that
the HERA-B 2003 $R(\chi_c)$ results~\cite{bib:chic_HERA-B_2003}
include a systematic uncertainty (of around 10\,\%) due to the
dependence of the detector's acceptance on the assumed J$/\psi$
polarization, taking into account that J$/\psi$'s from $\chi_c$ decays
may have a polarisation different from the directly produced ones.
This effect was not considered by the previous experiments, given
the poor statistical accuracy of their measurements.

Given the considerations expressed above, we have selected for our
analysis the $\rho(\psi^\prime)$ and $R(\chi_c)$ measurements listed
in Table~\ref{tab:selected}.

\begin{table}[h!]
\begin{center}
\begin{tabular}{|l|c|c|c|}
    \hline
    \footnotesize{Experiment} \hspace{20 pt} &
    \footnotesize{Target nucleus} & \footnotesize{$L$ [fm]}  &
    \hspace{30 pt} \footnotesize{$\rho(\psi^\prime)$ [\%]} \hspace{30 pt} \\
    \hline
    \hline
    NA51  &
    \begin{tabular}{@{}c@{}}  H  \\  D   \\ \end{tabular} &
    \begin{tabular}{@{}c@{}}  0  \\ 0.13 \\ \end{tabular} &
    \begin{tabular}{@{}c@{}}
    $1.57 \pm 0.05$ \\
    $1.67 \pm 0.06$ \\
    \end{tabular} \\

    \hline
    NA50 96/98  &
    \begin{tabular}{@{}c@{}} Be \\ Al \\ Cu \\ Ag \\ W  \\ \end{tabular} &
    \begin{tabular}{@{}c@{}} 0.86 \\ 1.84 \\ 2.66 \\ 3.41 \\ 3.93  \\ \end{tabular} &
    \begin{tabular}{@{}c@{}}
    $1.720 \pm 0.041$ \\
    $1.725 \pm 0.035$ \\
    $1.645 \pm 0.026$ \\
    $1.580 \pm 0.026$ \\
    $1.528 \pm 0.035$ \\
    \end{tabular} \\

    \hline
    NA50 2000  &
    \begin{tabular}{@{}c@{}} Be \\ Al \\ Cu \\ Ag \\ W \\ Pb \\ \end{tabular} &
    \begin{tabular}{@{}c@{}} 0.86 \\ 1.84 \\ 2.66 \\ 3.41 \\ 3.93 \\ 4.28 \\ \end{tabular} &
    \begin{tabular}{@{}c@{}}
    $1.745 \pm 0.086$ \\
    $1.889 \pm 0.079$ \\
    $1.593 \pm 0.082$ \\
    $1.599 \pm 0.085$ \\
    $1.422 \pm 0.079$ \\
    $1.461 \pm 0.068$ \\
    \end{tabular} \\

    \hline
    \hline

    \footnotesize{Experiment} \hspace{20 pt} &
    \multicolumn{2}{|c|}{ \footnotesize{$\langle x_{\mathrm{F}} \rangle$} }  &
    \footnotesize{$\rho(\psi^\prime)_{\mathrm{W}} ~/~ \rho(\psi^\prime)_{\mathrm{Be}}$} \\
    \hline
    \hline

    E866  &
    \multicolumn{2}{|c|}{
    \begin{tabular}{@{}c@{}}
    $-0.065$ \\
    $-0.019$ \\
    $~0.027$ \\
    $~0.075$ \\
    $~0.124$ \\
    $~0.173$ \\
    \end{tabular} } &

    \begin{tabular}{@{}c@{}}
    $0.904 \pm 0.068$ \\
    $0.900 \pm 0.038$ \\
    $0.932 \pm 0.030$ \\
    $0.932 \pm 0.031$ \\
    $0.939 \pm 0.036$ \\
    $0.881 \pm 0.048$ \\
    \end{tabular} \\

    \hline
    \hline


    \footnotesize{Experiment} \hspace{20 pt} &
    \footnotesize{Target nucleus} & \footnotesize{$L$ [fm]}  &
    \hspace{30 pt} \footnotesize{$R(\chi_c)$ [\%]} \hspace{30 pt} \\
    \hline
    \hline

    ISR & p & 0 & $35 \pm 6~$  \\

    \hline
    E705 & Li & 0.80 & $30 \pm 6~$  \\

    \hline
    HERA-B 2000 &
    \begin{tabular}{@{}c@{}} C \\ Ti \\  \end{tabular} &
    \begin{tabular}{@{}c@{}} 1.22 \\ 2.30 \\  \end{tabular} &
    \begin{tabular}{@{}c@{}} $36 \pm 10$ \\ $33 \pm 17$ \\ \end{tabular} \\

    \hline
    HERA-B 2003 &
    \begin{tabular}{@{}c@{}} C \\ W \\  \end{tabular} &
    \begin{tabular}{@{}c@{}} 1.22 \\ 3.93 \\  \end{tabular} &
    \begin{tabular}{@{}c@{}} $20.2 \pm 3.3~$ \\ $21.1 \pm 4.4~$ \\ \end{tabular} \\
    \hline

\end{tabular}
\caption{\label{tab:selected} The $\rho(\psi^\prime)$ and $R(\chi_c)$
      measurements selected for the present analysis.  The $L$ values
      correspond to an average nuclear density of $0.17$~fm$^{-3}$.}
\end{center}
\end{table}


\section{\boldmath J$/\psi$ feed-down from $\psi^\prime$ decays}
\label{sec:psiprime}

\begin{figure}[h!]
\centering \resizebox{0.6\textwidth}{!}{
\includegraphics*{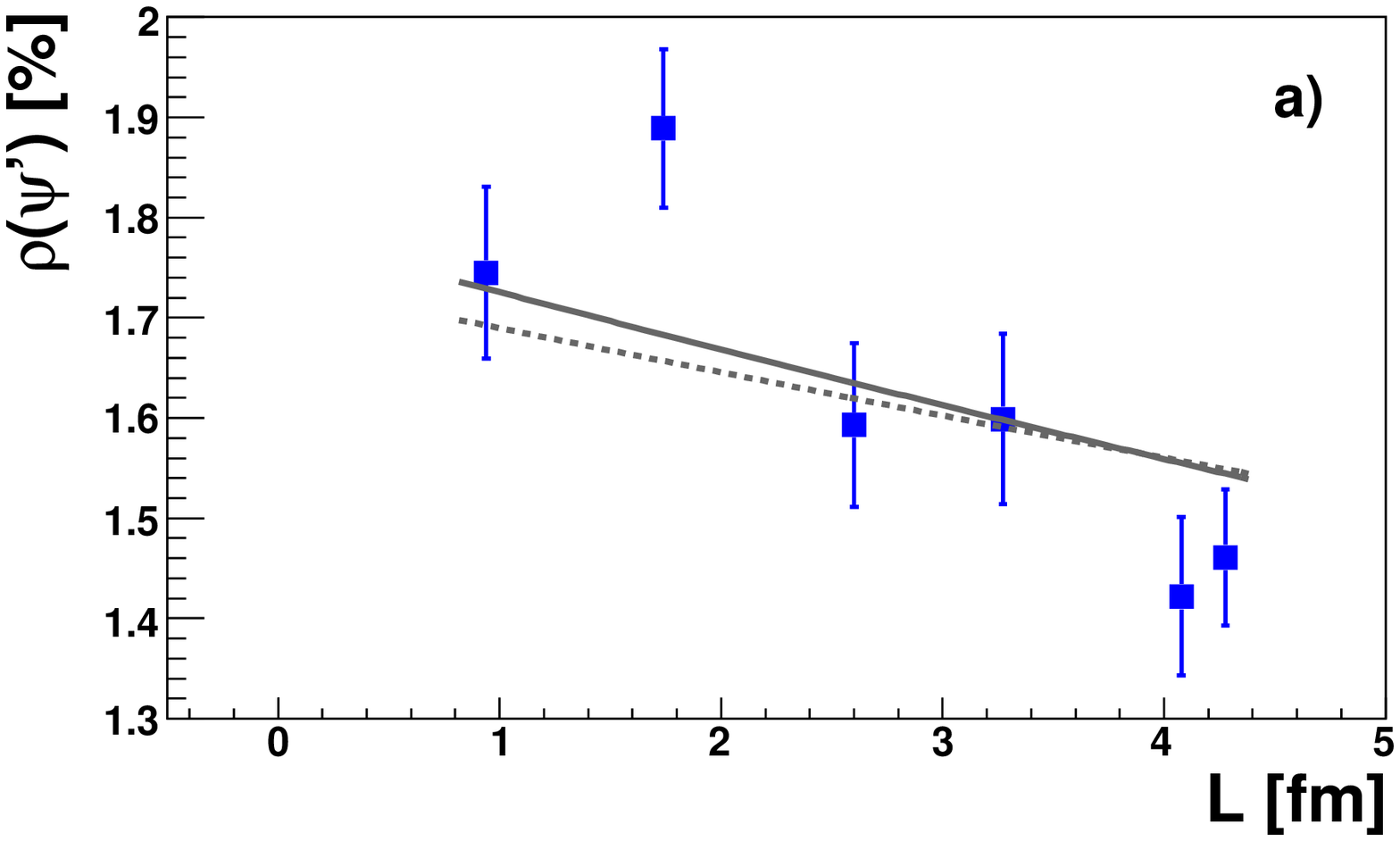}}
\centering \resizebox{0.6\textwidth}{!}{
\includegraphics*{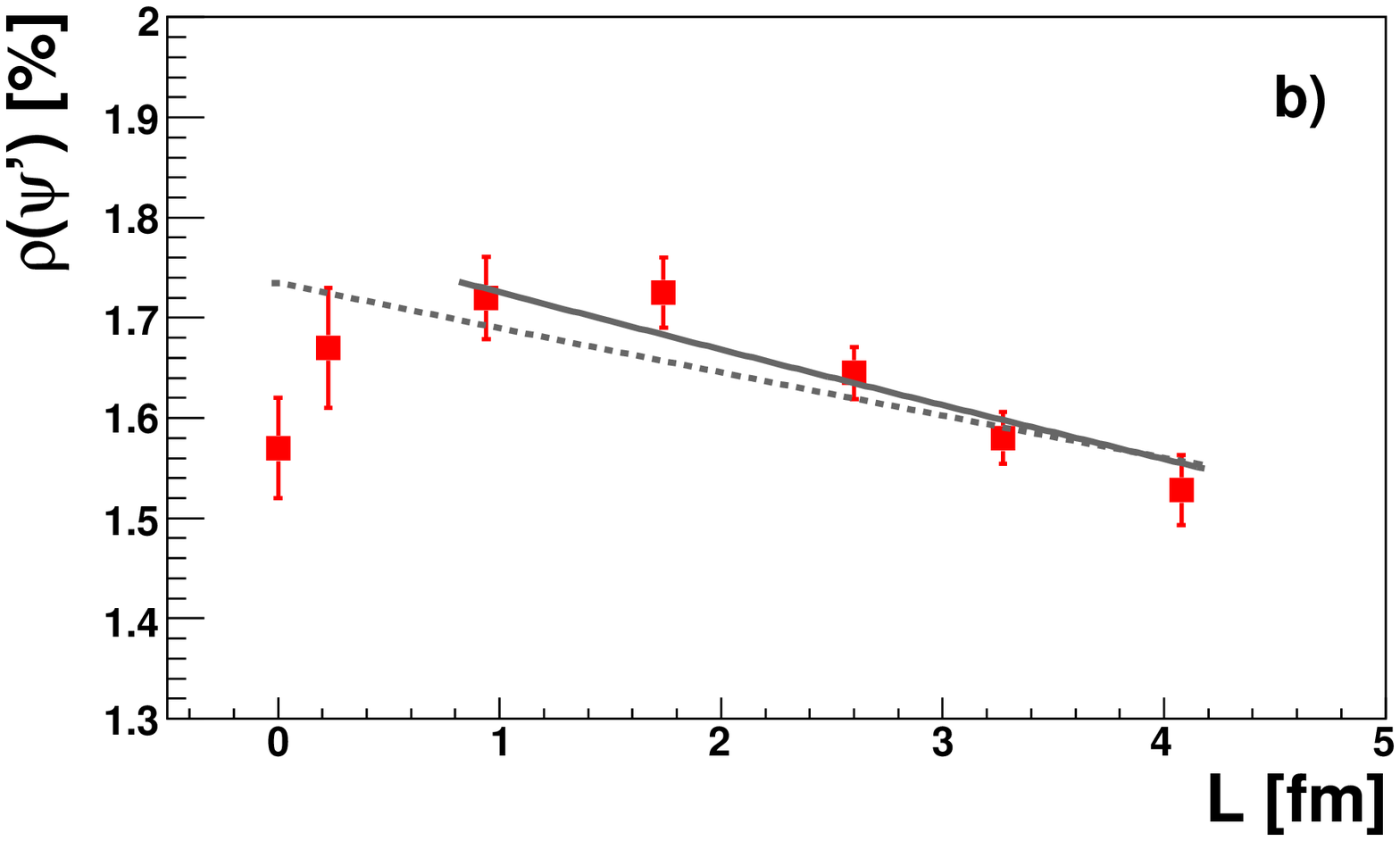}}
\centering \resizebox{0.6\textwidth}{!}{
\includegraphics*{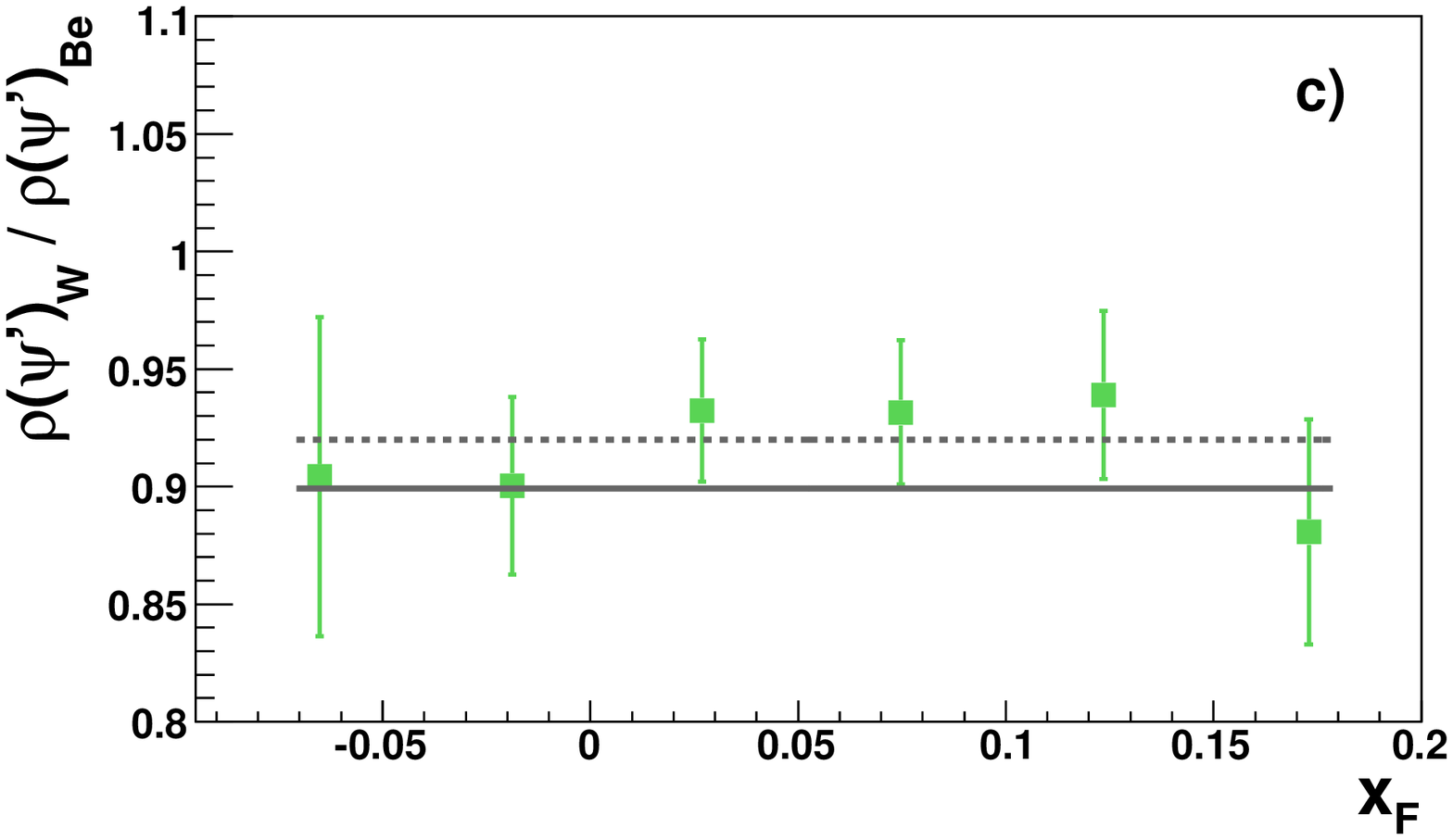}}
\caption{\label{fig:psip_points} The
    $\rho(\psi^\prime)$ values as a function of the nuclear path
    length, $L$, from NA50/51 measurements at 400~GeV (a) and 450~GeV
    (b), and the
    $\rho(\psi^\prime)_{\mathrm{W}}\,/\,\rho(\psi^\prime)_{\mathrm{Be}}$
    ratio measured by E866, in the $-0.1<x_{\mathrm{F}}<0.2$ window
    (c).  The curves are the result of the global fit described in the
    text, including (dashed lines) or excluding (solid lines) the NA51
    points.}
\end{figure}

The experimental points selected for the determination of the J$/\psi$
feed-down contribution from $\psi^\prime$ decays are shown in
Fig.~\ref{fig:psip_points}, as a function of the size of the target
nucleus (in the case of the SPS data) or of $x_{\mathrm{F}}$ (in the
case of E866).  These measurements clearly show that the $\psi^\prime$
and J$/\psi$ states are differently absorbed by the nuclear medium.

In order to determine the $\psi^\prime$-to-J$/\psi$ feed-down fraction
in pp collisions, $R^0(\psi^\prime)$, all selected measurements were
simultaneously fitted within the framework of the Glauber
formalism~\cite{bib:rhoL}, using the so-called ``$\rho L$
parametrization'':
\begin{equation}
\sigma(pA \rightarrow \psi) ~/~ A\,\sigma(pN \rightarrow \psi) \, = \, \exp(-
\sigma_{\mathrm{abs}} \, \rho \, L) \quad , \nonumber
\end{equation}
where $\rho$ is the nuclear density and $L$ is the nuclear
path length traversed by the charmonium state, of absorption cross section
$\sigma_{\mathrm{abs}}$. The $\rho\, L$ values were determined through a Glauber calculation,
for each nuclear target, taking into account the appropriate nuclear density profiles, as
described in Ref.~\cite{bib:psip_NA50_400GeV}. The fit provides two parameters: the
$R^0(\psi^\prime)$ ``reference'' feed-down fraction (corresponding to $L = 0$) and the
difference between the $\psi^\prime$ and J$/\psi$ absorption cross sections, where the
J$/\psi$ term does not include the $\psi^\prime$ decay contribution (to remove
auto-correlation effects).

It should be kept in mind that this parametrization represents a
rather simplified description of the nuclear absorption process,
convoluting in a single \emph{effective} absorption cross section a
multitude of physical effects.  In particular, $\sigma_{\mathrm{abs}}$
is assumed to be a ``universal quantity'', independent of the
collision energy and of the kinematical properties of the produced
charmonium states.  It also implicitly incorporates the nuclear modifications
of the parton distribution functions, possible energy loss mechanisms,
formation time effects, etc.  Following most previous studies of
charmonium absorption in nuclear matter, we use this simple
parametrization in the analysis presented in this paper, where we
focus on the mid-rapidity results; this issue will be revisited in the
future, in the scope of a broader investigation.

\begin{figure}[h!]
\centering \resizebox{0.6\textwidth}{!}{
\includegraphics*{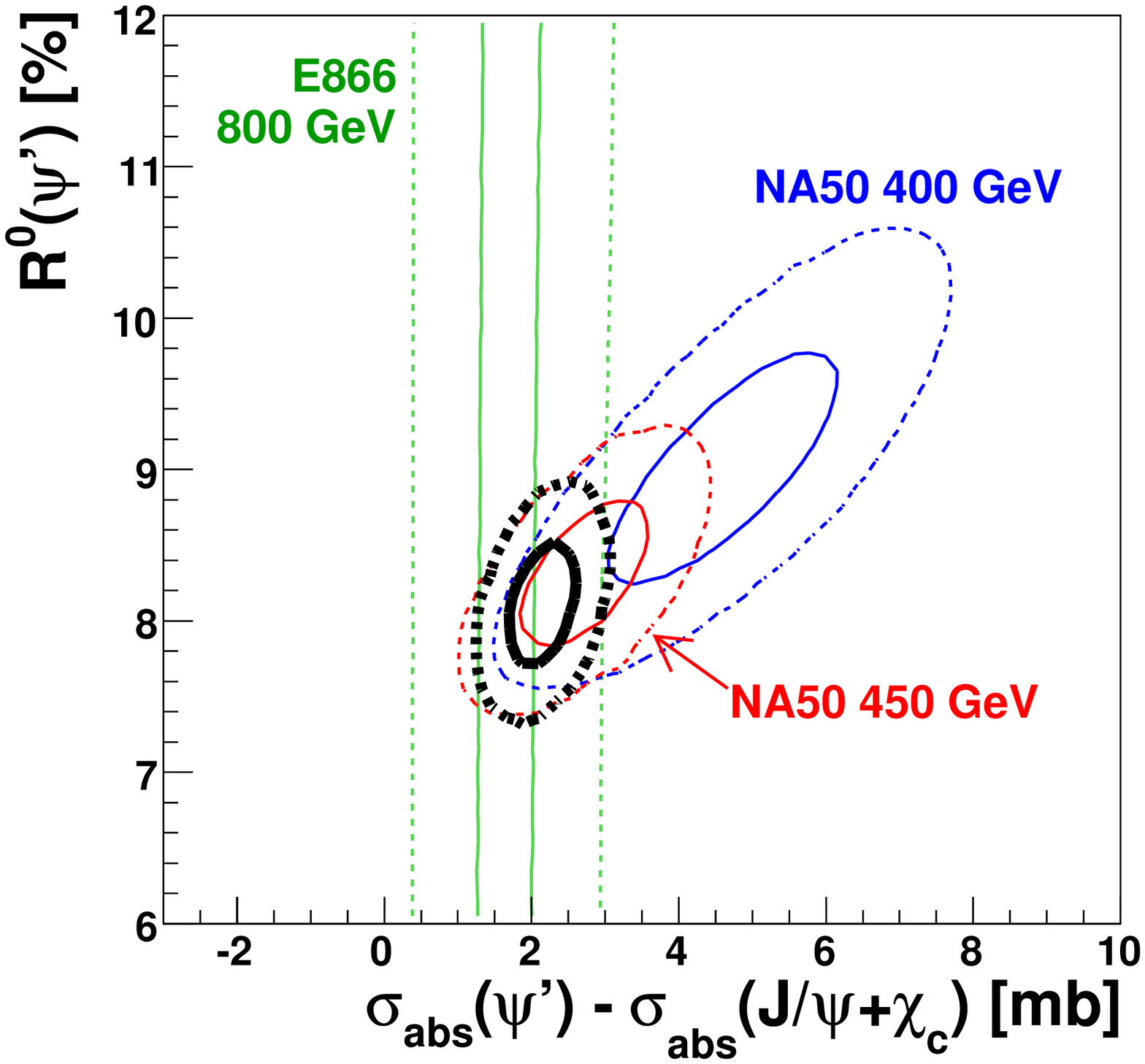}} \caption{ \label{fig:psip_contours} $68\,\%$ (solid
    lines) and
    $99\,\%$ (dashed lines) confidence level contours for the bi-dimensional
    probability distribution of the parameters $R^0(\psi^\prime)$ and
    $\sigma_{\mathrm{abs}}(\psi^\prime)-\sigma_{\mathrm{abs}}(\mathrm{J}/\psi+\chi_c)$.
    The thick black contours delimit the region favoured by the global fit,
    while the thin coloured (or grey) ones reflect the individual data sets.}
\end{figure}

A global fit to all data points leads to the dashed lines in
Fig.~\ref{fig:psip_points}, with a chi-square probability of only
$1\,\%$, clearly indicating that the model is unable to properly account
for the NA51 measurements, performed with hydrogen and deuterium
targets (the two leftmost points in the middle panel).  Maybe
the fact that protons and deuterons are exceptionally light nuclei places
them out of the domain of applicability of the model we are using because
they are not large enough to be traversed by fully formed charmonium
states.
It should also be noted that the use of ``nuclear density profiles'' in
the Glauber calculation of the proton and deuteron $\rho \, L$ values
is not as reliable as in the case of the heavier nuclei.  Furthermore,
it is not clear that the same value, 0.17~nucleon/fm$^3$, should be
used as average nuclear density for all nuclei, including protons and
deuterons, when extracting $L$ from the calculated $\rho \, L$ values.
Without the pp and p-D points, the best description of the data is
represented by the solid lines, with a chi-square probability of
$27\,\%$, reflecting a much better compatibility between the data and
the model used in our fit.  The corresponding feed-down fraction is
\begin{equation}
R^0(\psi^\prime) = (8.1 \pm 0.3) \,\% \quad .
\end{equation}
Including the NA51 points decreases the result to $(7.9 \pm 0.3)\,\%$,
a negligible change despite the visible degradation of the fit
quality.

The correlation between the two fit parameters is shown in
Fig.~\ref{fig:psip_contours}, as a bi-dimensional contour plot.
Although the three data sets give compatible results, they
nevertheless indicate that the difference between the charmonium
absorption cross sections decreases with increasing collision energy.

\section{\boldmath J$/\psi$ feed-down from $\chi_c$ decays}
\label{sec:chic}

\begin{figure}[htb]
\centering \resizebox{0.6\textwidth}{!}{
\includegraphics*{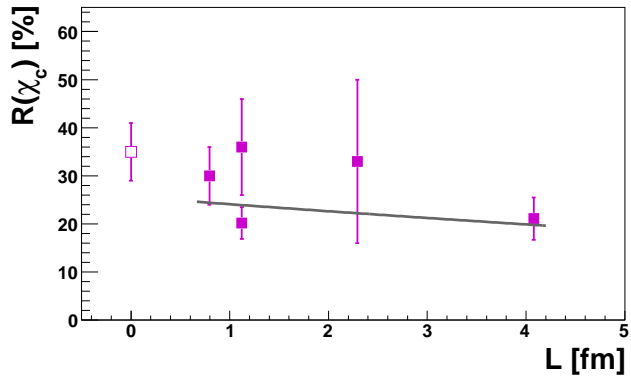}} \caption{ \label{fig:chic_points} The $R(\chi_c)$
    measurements used in the present analysis as a function of the
    nuclear path length $L$. The curve is the result of the fit
    described in the text (excluding the first point).}
\end{figure}

\begin{figure}[htb]
\centering \resizebox{0.6\textwidth}{!}{
\includegraphics*{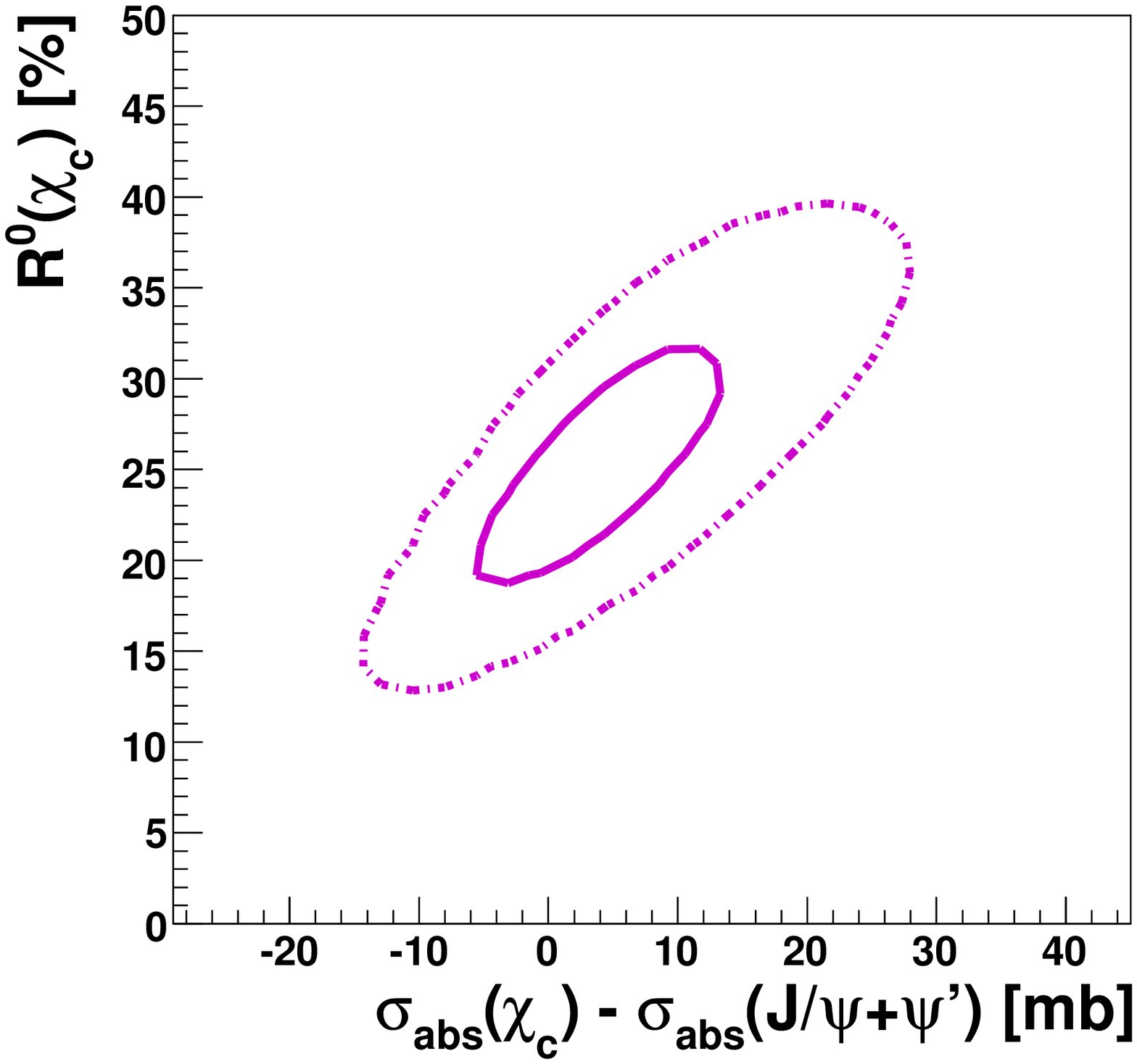}} \caption{ \label{fig:chic_contours} $68\,\%$ and
    $99\,\%$ confidence level contours for the bi-dimensional
    probability distribution of the fit parameters $R^0(\chi_c)$ and
    $\sigma_{\mathrm{abs}}(\chi_c)-\sigma_{\mathrm{abs}}(\mathrm{J}/\psi+\psi^\prime)$.}
\end{figure}

The $R(\chi_c)$ values collected in Table~\ref{tab:selected} are shown
in Fig.~\ref{fig:chic_points} as a function of $L$.  The curve is the
result of a fit analogous to the one explained in the previous
section, using the ``$\rho L$ parametrization'' and leaving free the
difference between the effective absorption cross sections of the
$\chi_c$ mesons and of the J$/\psi$ mesons not coming from $\chi_c$
decays. Given the conjecture, suggested by the $\psi^\prime$ analysis,
that measurements performed with very light nuclei are not accountable
within the simple absorption model adopted here, the pp point is
excluded from the fit. The resulting feed-down fraction (for $L = 0$)
is
\begin{equation}
R^0(\chi_c) = (25 \pm 5) \,\% \quad ,
\end{equation}
with a fit $\chi^2$ probability of $25\,\%$.

The $R^0(\chi_c)$ value considerably depends on the difference between
the absorption cross sections of the two charmonium states (see
Fig.~\ref{fig:chic_contours}).  Therefore, a more precise
$R^0(\chi_c)$ value can be obtained if an improved understanding of
charmonium absorption in nuclear targets significantly reduces the
allowed range of $\sigma_{\mathrm{abs}}(\chi_c) -
\sigma_{\mathrm{abs}}({\rm J}/\psi)$.

\section{Summary}

We presented and reviewed the presently available $\psi^\prime$ and
$\chi_c$ hadro-production measurements, and derived global averages of
the J$/\psi$ feed-down fractions from $\psi^\prime$ and $\chi_c$
decays, at mid-rapidity:
\begin{equation}
  R^0(\psi^\prime) = (8.1 \pm 0.3) \,\% \quad, \quad
  R^0(\chi_c) = (25 \pm 5) \,\% \quad .
\end{equation}
These averages reflect measurements performed at collision energies up
to $\sqrt{s} \sim 60$~GeV.  At much higher energies, CDF measured
$R(\chi_c) = (30 \pm 7)\,\%$ in $\mathrm{p}\bar{\mathrm{p}}$
collisions at $\sqrt{s}=1800$~GeV~\cite{bib:chic_CDF} and PHENIX
reported preliminary values obtained in pp collisions at
$\sqrt{s}=200$~GeV: $R(\psi^\prime) = (8.6 \pm 2.5)\,\%$ and
$R(\chi_c) < 42\,\%$ (at $90\,\%$ C.L.)~\cite{bib:PHENIX}.
More precise measurements would be needed to probe an eventual energy
dependence of the J$/\psi$ feed-down fractions from decays of heavier
charmonium states.


Since most of the existing measurements were performed with nuclear
targets, the derivation of the $R^0(\psi^\prime)$ and $R^0(\chi_c)$
values relevant for elementary collisions ($L=0$) requires modelling
the influence of the nuclei on the production yields of the three
charmonium states.  In the study reported in this paper we followed a
widespread model where the three charmonium states are analogously
absorbed while traversing the nuclear matter, with three (a priori
different) effective absorption rates.
This rather simple model provides a reasonable description of the
available measurements if we restrict the analysis to the mid-rapidity
data and exclude values obtained with ``exceptionally light nuclei''
(protons and deuterons).  The extension of our analysis to a broader
set of measurements requires an improved phenomenological model, which
should reflect the following observations.
It is only for nuclear targets heavier than beryllium that the
J$/\psi$ and $\psi^\prime$ nuclear absorption rates are significantly
different.  This difference decreases when the collision energy
increases and when we approach forward $x_{\mathrm{F}}$.
The $R(\chi_c)$ value derived from data collected with heavy nuclei
and at small $|x_{\mathrm{F}}|$, $0.22 \pm 0.03$, is significantly
smaller than the value derived from data collected with light nuclei
and at forward $x_{\mathrm{F}}$, $0.36 \pm 0.02$.

These might be indications that a proper understanding of charmonium
absorption requires considering that the objects traversing the
nuclear matter are not fully formed J$/\psi$, $\psi^\prime$ or
$\chi_c$ states but rather pre-resonance states having a suitable time
evolution.  This is the topic of a more complex investigation, to be
reported in a future publication.

\bigskip

We would like to acknowledge very useful discussions with Ramona Vogt.
This work was partially supported by the Funda\c{c}\~ao para a
Ci\^encia e a Tecnologia, Portugal, under contracts
SFRH/BPD/42343/2007, SFRH/BPD/42138/2007 and CERN/FP/83516/2008.


\begin{thebibliography}{999}

\bibitem{bib:Karsch}
F. Karsch, ``Lattice QCD at High Temperature and Density'', in ``Lectures on Quark
Matter'', W. Plessas and L. Mathelitsch (eds.), Springer (2002), p. 209.

\bibitem{bib:MatsuiSatzKarschMehr}
T. Matsui and H. Satz, \plb{178}{1986}{416}; \\
F. Karsch, M.-T. Mehr and H. Satz, \zpc{37}{1988}{617}.

\bibitem{bib:KharzeevKarschSatz} F. Karsch, D. Kharzeev and H. Satz, \plb{637}{2006}{75}.

\bibitem{bib:PDG} C. Amsler et al. (Particle Data Group), \plb{667}{2008}{1}.

\bibitem{bib:psip_E331}
J.G. Branson et al. (E331 Coll.), \prl{38}{1977}{1331}.

\bibitem{bib:psip_E444}
K.J. Anderson et al. (E444 Coll.), \prl{42}{1979}{944}.

\bibitem{bib:psip_E705}
L. Antoniazzi et al. (E705 Coll.), \prd{46}{1992}{4828}.

\bibitem{bib:psip_E288}
H.D. Snyder et al. (E288 Coll.), \prl{36}{1976}{1415}.

\bibitem{bib:psip_NA38}
M.C. Abreu et al. (NA38 Coll.), \plb{444}{1998}{516}; \\
C. Louren\c{c}o, PhD thesis, Universidade T\'{e}cnica de Lisboa, 1995.

\bibitem{bib:psip_NA51}
M.C. Abreu et al. (NA51 Coll.), \plb{438}{1998}{35}.

\bibitem{bib:psip_NA50_450GeV}
B. Alessandro et al. (NA50 Coll.), \epjc{33}{2004}{31}.

\bibitem{bib:psip_NA50_400GeV}
B. Alessandro et al. (NA50 Coll.), \epjc{48}{2006}{329}.

\bibitem{bib:psip_E771}
T. Alexopoulos et al. (E771 Coll.), \plb{374}{1996}{271}.

\bibitem{bib:psip_E789}
M.H. Schub et al. (E789 Coll.), \prd{52}{1995}{1307}.

\bibitem{bib:psip_E866}
M.J. Leitch et al. (E866 Coll.), \prl{84}{2000}{3256} \\ and
\texttt{http://p25ext.lanl.gov/e866/papers/e866prlj/ratiosc.txt}.

\bibitem{bib:psip_HERA-B}
I. Abt et al. (HERA-B Coll.), \epjc{49}{2007}{545}.

\bibitem{bib:psip_WA39}
M.J. Corden et al. (WA39 Coll.), \plb{96}{1980}{411}.

\bibitem{bib:psip_E537}
C. Akerlof et al. (E537 Coll.), \prd{48}{1993}{5067}.

\bibitem{bib:psip_WA11}
M.A. Abolins et al. (WA11 Coll.), \plb{82}{1979}{145}.

\bibitem{bib:psip_E615}
J.G. Heinrich et al. (E615 Coll.), \prd{44}{1991}{1909}.

\bibitem{bib:psip_E672}
A. Gribushin et al. (E672-706 Coll.), \prd{53}{1996}{4723}.

\bibitem{bib:psip_ISR}
A.G. Clark et al. (R702 Coll.), \npb{142}{1978}{29}.

\bibitem{bib:chic_E369}
T.B.W. Kirk et al. (E369-610-673 Coll.), \prl{42}{1979}{619}; \\
S.R. Hahn et al. (E369 Coll.), \prd{30}{1984}{671}; \\
D.A. Bauer et al (E369-610-673 Coll.), \prl{54}{1985}{753}.

\bibitem{bib:chic_E705}
L. Antoniazzi et al. (E705 Coll.), \prl{70}{1993}{383}.

\bibitem{bib:chic_E771}
T. Alexopoulos et al. (E771 Coll.), \prd{62}{2000}{032006}.

\bibitem{bib:chic_HERA-B_2000}
I. Abt et al. (HERA-B Coll.), \plb{561}{2003}{61}.

\bibitem{bib:chic_HERA-B_2003}
P. Faccioli for the HERA-B Coll., Int. Workshop on
Heavy Quarkonium, DESY, Hamburg, October 2007.

\bibitem{bib:chic_SERPUKHOV-140}
F. Binon et al. (SERPUKHOV-140 Coll.), \npb{239}{1984}{311}.
xs
\bibitem{bib:chic_WA11}
Y. Lemoigne et al. (WA11 Coll.), \plb{113}{1982}{509}.

\bibitem{bib:chic_E672}
V. Koreshev et al. (E672-706 Coll.), \prl{77}{1996}{4294}.

\bibitem{bib:chic_ISR}
J.H. Cobb et al., \plb{72}{1978}{497}, \\
C. Kourkoumelis et al., \plb{81}{1979}{405}, \\
and Ref.~\cite{bib:psip_ISR}.

\bibitem{bib:chic_CDF}
F. Abe et al. (CDF Coll.), \prl{79}{2003}{578}.

\bibitem{bib:VogtXF}
R. Vogt, \prc{61}{2000}{035203}.

\bibitem{bib:rhoL} D. Kharzeev, C. Louren\c{c}o, M. Nardi and H. Satz,
\zpc{74}{1997}{307}.

\bibitem{bib:PHENIX}
E.T. Atomssa for the PHENIX Coll., 3rd Int.
Conf. on Hard and Electromagnetic Probes of High-Energy Nuclear Collisions, A Toxa,
Spain, June 2008.

\end{thebibliography}
\end{document}